\documentclass[aps,prd,eqsecnum,11pt,showpacs,nofootinbib]{revtex4-1}
\usepackage{geometry}                		
\usepackage{graphicx}				
\usepackage{amssymb}
\usepackage{graphicx}
\usepackage{amssymb,amsmath}
\usepackage{amsfonts,mathrsfs}

\newcommand{\be}{\begin{equation}}
\newcommand{\ee}{\end{equation}}
\newcommand{\bea}{\begin{eqnarray}}
\newcommand{\eea}{\end{eqnarray}}
\newcommand{\bes}{\begin{subequations}}
\newcommand{\ees}{\end{subequations}}

\begin{document}

\title{ Scattering coefficients and gray-body factor for \\ 1D BEC acoustic black holes: exact results}

\author{Alessandro~Fabbri}
\email{afabbri@ific.uv.es}
\affiliation{Centro Studi e Ricerche E. Fermi, Piazza del Viminale 1, 00184 Roma, Italy}
\affiliation{Dipartimento di Fisica dell'Universit\`a di Bologna and INFN sezione di Bologna, Via Irnerio 46, 40126 Bologna, Italy}
\affiliation{Laboratoire de Physique Th\'eorique, CNRS UMR 8627, B\^at. 210, Universit\'e Paris-Sud 11, Univ. Paris-Saclay, 91405 Orsay Cedex, France}
\affiliation{Departamento de F\'isica Te\'orica and IFIC, Universidad de Valencia-CSIC, C. Dr. Moliner 50, 46100 Burjassot, Spain}
\author{Roberto~Balbinot}
\email{Roberto.Balbinot@bo.infn.it}
\affiliation{Dipartimento di Fisica dell'Universit\`a di Bologna and INFN sezione di Bologna, Via Irnerio 46, 40126 Bologna, Italy}
\affiliation{Centro Studi e Ricerche E. Fermi, Piazza del Viminale 1, 00184 Roma, Italy}
\author{Paul~R.~Anderson}
\email{anderson@wfu.edu}
\affiliation{Department of Physics, Wake Forest University, Winston-Salem, North Carolina 27109, USA}

\begin{abstract}
  A complete set of exact analytic solutions to the mode equation is found in the region exterior to the acoustic horizon
for a class of 1D Bose-Einstein condensate (BEC) acoustic black holes.
 From these, analytic expressions for the scattering coefficients and gray-body factor are obtained. The results are used
 to verify previous predictions regarding the behaviors of the scattering coefficients and gray-body factor in the low frequency limit.
 \end{abstract}

\maketitle

\section{Introduction}

The particles emission by black holes, discovered by Hawking in 1974~\cite{Hawking:1974sw}, does not have an exact Planckian distribution.  Instead this distribution is modified
by a so-called gray-body factor $\Gamma_i(\omega)$ such that
\be N_\omega^{(i)}=\frac{\Gamma_i(\omega)}{e^{\frac{\hbar\omega}{k_B T_H}}-1}\ .\ee
Here  $N_{\omega}^{(i)}$ is the number of quanta emitted of energy $\omega$ and additional quantum numbers labelled by ``$i$'', and $T_H$ is the Hawking temperature which is proportional to the surface gravity of the black hole (BH) event horizon.

Spacetime curvature around the BH induces in the
mode equation an effective potential that causes backscattering of the modes. Because of this a mode originating from the horizon region is only partially transmitted to future infinity. $\Gamma^i(\omega)$ measures the probability for this to happen.
The calculation of the gray-body factor, being a nonlocal effect, requires the exact form of the modes to be known.  Due to the mathematical complexity of the mode
equation, this seldom happens.  Thus only in certain cases, such as the low frequency limit~\cite{a,b,c,d,e,abfp2,afb} or when the
effective potential vanishes and propagation is trivial ($\Gamma=1$), have analytic expressions for the gray-body factor previously been obtained.

We have found a nontrivial example in which analytic solutions to the mode equation can be found and an analytic expression for the gray-body factor can be obtained in a rather interesting physical situation that can, in principle, be reproduced in an experimental setting.
It involves a model for a Bose-Einstein condensate, BEC, which is arranged to serve as an analog model for a black hole.
Similar models have
been investigated previously using either analytic approximations or numerical computations~\cite{bffrc,cfrba,rpc,pama,mrfpbc,lrcp,abfp1}.
The condensate is approximated as being infinite in length and can be treated as being effectively
 one-dimensional.  It moves at a constant velocity in the lab frame.
The system is set up so that the speed of sound in the condensate varies as a function of position in the direction of the flow such that there is one region for which the flow is supersonic and one where it is subsonic.  At one point (the acoustic horizon) the speed of sound is equal to the flow speed.
What is different about our model is that we have found a speed of sound profile for which nontrivial analytic solutions to the mode equation can be obtained.

The primary reason for studying analog systems is that experimental detection of Hawking radiation in an astrophysical context is unlikely. Indeed, for a black hole formed from gravitational collapse $T_H\leq 10^{-7}\ K$ is several orders of magnitudes smaller than the cosmic microwave background temperature $T_{CMB}\sim 3\ K$.   Fortunately, Hawking radiation is not specific to gravity. In 1981 Unruh \cite{unruh} used the mathematical equivalence between the propagation
of a scalar field in a curved spacetime background and that of sound in an inhomogeneous eulerian fluid (the so called gravitational analogy \cite{blv})
to show that fluids undergoing subsonic-supersonic transition (acoustic black holes) will emit analog Hawking radiation in the form of phonons due to the presence of an acoustic horizon.
This has opened up
new possibilities to either directly detect Hawking radiation experimentally or to detect certain effects associated with Hawking radiation.  Experiments have been done with water tanks \cite{wt}, quantum optics \cite{qo},  polaritons \cite{po}, and Bose-Einstein condensates (BECs)  \cite{bec}. A major difficulty is that the signal for Hawking radiation is weak in
most systems compared to background effects.  In this regard, BECs offer particularly favorable experimental conditions since it is possible to have a background temperature as
low as $100 nK$ which is only an order of magnitude larger than the Hawking temperature of $10K$.

In Section II we review the mathematical details for this type of model, focusing on the form of the mode equation and its asymptotic solutions.
In Section III we use our profile to find a complete set of exact analytic solutions to the mode equation.  From these solutions we obtain analytic expressions for the scattering coefficients and the gray-body factor.  Section IV contains a brief discussion of our results.

\section{BEC analog black hole model}

To make contact with the gravitational analogy \cite{blv}, one considers the hydrodynamic approximation of BEC theory. As usual ~\cite{pit-str}, we write down the bosonic field operator $\hat\Psi$ in the density - phase representation $\hat\Psi=\sqrt{\hat n}e^{i\hat\theta}$
and expand the density operator $\hat n = n+\hat n_1$ and the phase operator  $\hat \theta = \theta + \hat \theta_1$ around a classical background (the condensate) described by the Gross-Pitaevski equation.
The small (quantum) fluctuations satisfy the Bogoliubov-de Gennes equations (see for instance \cite{book})
\bea \hbar \partial_t\hat\theta_1 &=&-\hbar \vec v \vec  \nabla \hat  \theta_1 -\frac{mc^2}{n}\hat n_1 +\frac{mc^2}{4n}\xi^2 \vec\nabla [n\vec\nabla(\frac{\hat n_1}{n})]=0\ , \label{bdg1}\\
\partial_t\hat n_1 &=& -\vec \nabla( \vec v \hat n_1 + \frac{\hbar n}{m}\vec\nabla \theta_1)\ , \label{bdg2} \eea
where $\vec v=\frac{\hbar\vec\nabla \theta}{m}$ is the condensate velocity, $c=\frac{gn}{m}$ the speed of sound, $\xi=\frac{\hbar}{mc}$ the healing length (the fundamental length scale of the system) and $m$ the mass of an individual atom. For backgrounds varying on length scales larger than $\xi$ the last term in (\ref{bdg1}) can be neglected and we recover the hydrodynamic approximation. We can then extract $\hat n_1$ from (\ref{bdg1}), i.e.
\be \label{hydron1} \hat n_1= -\frac{\hbar n}{mc^2}[\vec v\vec\nabla \hat \theta_1 +\partial_t\hat \theta_1]\ .\ee
Substituting~\eqref{hydron1} into (\ref{bdg2})
one finds that the phase fluctuation $\hat \theta_1$ obeys
\be\label{u} -(\partial_t + \vec \nabla \vec v)\frac{n}{mc^2}(\partial_t + \vec v \vec \nabla )\hat\theta_1+\vec\nabla \frac{n}{m}\vec\nabla \hat \theta_1 =0\ , \ee
which is formally equivalent to the Klein-Gordon equation
\be \label{kg} \Box \hat \theta_1 = \frac{1}{\sqrt{-g}}\partial_\mu (\sqrt{-g}g^{\mu\nu}\partial_\nu \hat \theta_1 )=0 \ee
in a fictitious curved space-time described by the acoustic metric
\be ds^2=\frac{n}{mc} \left[-(c^2-\vec v^2)dt^2 - 2\vec v d\vec x dt + d\vec x^2 \right] \;. \ee

We consider a model in which  a BEC moves at a constant velocity $\vec{v} = - v_0 \hat{x}$ with $v_0 >0$, and the
number density $n$ of the atoms is constant.
Using Feshbach resonances~\cite{pit-str} it is possible to have a variable speed of sound $c$.
The speed of sound is arranged so that it varies only in the direction of the flow with $c(x)>v_0$ for $x>0$ and
 $c(x) <v_0$ for $x<0$. The latter region represents the acoustic black hole and $x=0$ is its horizon.
   The condensate is also asymptotically homogeneous so that $c(x)\to c_R$ for $x\to +\infty$ and $c(x)\to c_L$ for $x\to -\infty$.
    Thus this system serves as an analog to a static asymptotically flat black hole.

We have seen that in the hydrodynamical approximation the phase fluctuation $\theta_1$ of the condensate satisfies (\ref{kg}) and thus behaves as a massless minimally coupled scalar field $\phi$
in the acoustic metric
\be ds^2=\frac{n}{mc} \left[-(c^2-v_0^2)dt^2 + 2v_0 dx dt + dx^2 + dx_{\perp}^2\right] \; , \ee
where $dx_\perp^2=dy^2+dz^2$.
It is useful to introduce a ``Schwarzschild time'' $t_S$ (as opposed to the laboratory ``Painlev\' e'' time $t$) such that
\be t_S=t - \int dx \frac{v_0}{c^2-v_0^2} \;. \ee
Then the acoustic metric becomes
\be \label{am} ds^2=\frac{n}{m} \left[ -\frac{c^2-v_0^2}{c}dt_S^2 + \frac{c}{c^2-v_0^2}dx^2 + \frac{dx_{\perp}^2}{c} \right] \;.  \ee

 If the phase fluctuations are quantized then the quantum field $\hat \phi\equiv \hat \theta_1$ can be expanded in terms of a complete set of mode functions $\varphi$ which are solutions to Eq.~\eqref{kg}.
The condensate is assumed to be infinite in extent in the flow ($x$) direction,
and that the  dynamics is frozen in the transverse direction.
We consider the so called ``1D mean field regime''  which assumes perfect 1D condensation, and is valid in an intermediate density regime \cite{pit-str, stringari, fpw}.
Restricting our attention to these modes and using separation of variables we can write
$\varphi(t,x)=e^{-i\omega t_S}\varphi_\omega (x)$.  Then Eq. (\ref{kg}) becomes
\be \label{fe} \frac{\omega^2}{c^2-v_0^2}\varphi + \frac{d}{dx} \left[ \frac{c^2-v_0^2}{c^2} \frac{d \varphi}{dx} \right] =0 \ . \ee

\begin{figure}[h]
\centering \includegraphics[angle=0, height=2.6in] {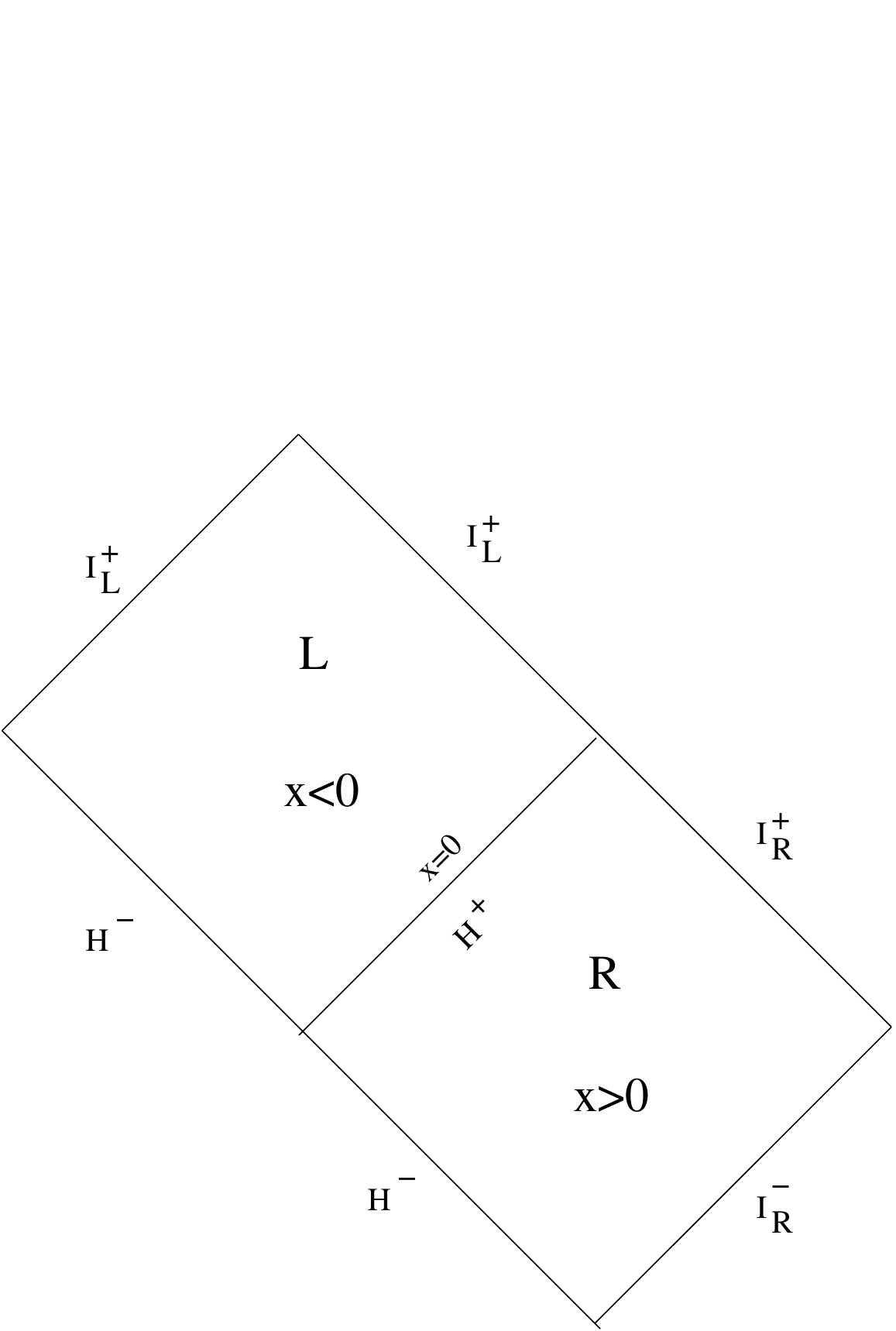}
\caption{Penrose diagram for our acoustic black hole. $I_R$ and $I_L$ are, respectively, the exterior and interior asymptotic regions. $H^+$ and $H^-$ are the future and past horizons.}
\label{penrose}
\end{figure}

Next we introduce a new space variable $z$ by the relation
\be \label{tr} \frac{dx}{(1-\frac{v_0^2}{c^2})} = 
 dz \;. \ee
The mode equation (\ref{fe}) becomes
\be \label{ff} \frac{\omega^2}{c^2(z)}\varphi + \frac{d^2 \varphi}{dz^2}  =0   \;. \ee
Note that this equation for $\varphi_\omega(z)$ has the same form as Eq.~\eqref{fe} for $\varphi_\omega(x)$ with $v_0$ set equal to zero.
Thus we have mapped the original problem in $x$ with $v_0 > 0$ to one in $z$ with $v_0 = 0$.
At the horizon $c = v_0$ so $z$ diverges.  Therefore the exterior ($x\ge 0$) and interior ($x\le 0$) regions must be dealt with separately.
A Penrose diagram representing the causal structure of the acoustic space-time under consideration is given in Fig.~\ref{penrose}.

 From Fig.~\ref{penrose} it is clear that for the $R$ region ($x\ge 0$), the union of the past horizon $H^-$ and past null infinity $I^-$ forms
 a Cauchy surface.  For each value of $\omega$ there are two linearly independent solutions to Eq.~(\ref{ff}), $\varphi_H$ and $\varphi_I$.
 The mode $e^{-i \omega t_S} \varphi_H$ vanishes on $I^-$ and is normalized on $H^-$, while the mode $e^{-i \omega t_s} \chi_I$ vanishes on $H^-$ and is normalized
 on $I^-$.  Together $e^{-i \omega t_S} \varphi_H(x) $ and $e^{-i \omega t_S} \varphi_I(x) $ form a complete set of solutions to the mode equation
in the $R$ region and thus can be used for the expansion of the quantum field $\phi$ in terms of modes.

The asymptotic behaviors of $\varphi_H$ are
 \begin{equation}\label{uu}
\varphi_H=\Big\{
  \begin{array}{c}
  \sqrt{\frac{v_0}{4\pi\omega}}  \left( e^{i \frac{\omega}{v_0} z}+R_H  e^{-i \frac{\omega}{v_0} z}\right)\ , \ \ x\to 0^+ \\
  \sqrt{\frac{c_R}{4\pi\omega}} \  T_H e^{i \frac{\omega}{c_R}z} , \ \  x \to +\infty \;. \\
   \end{array}
   \end{equation}
Multiplying by $e^{-i\omega t_S}$ one sees that this solution corresponds to a unit norm outgoing wave originating on $H^-$ which is partially reflected back to the future horizon $H^+$ and partially transmitted to future null infinity $I^+$. Here $R_H$ and $T_H$ are the reflection and transmission coefficient respectively for this mode. The gray-body factor is $\Gamma(\omega)=|T_H|^2$.
The asymptotic behaviors of $\varphi_I$ are
 \begin{equation}\label{II}
\varphi_I=\Big\{
  \begin{array}{c}
  \sqrt{\frac{v_0}{4\pi\omega}}\  T_I e^{-i \frac{\omega}{v_0}z} \ , \ \ x\to 0^+ \\
  \sqrt{\frac{c_R}{4\pi\omega}} \  \left( e^{-i \frac{\omega}{c_R} z}+R_I  e^{i \frac{\omega}{c_R} z}\right) . \ \  x \to +\infty \\
   \end{array}
   \end{equation}
   Multiplying by $e^{-i \omega t_S}$ one sees that this corresponds to a unit norm incoming wave originating on $I_R^-$  which is partially reflected back to infinity ($I_R^+$) with a reflection
   coefficient $R_I$, and which is partially transmitted towards the horizon with a transmission coefficient $T_I$.
   The factors $\sqrt{v_0}$ and $\sqrt{c_R}$ appearing in $\varphi_H$ and $\varphi_I$ come from the proper normalization of the modes along the corresponding Cauchy surfaces and are consequences of the $\frac{1}{c}$ factor present in the metric (\ref{am})
   in the transverse part.

   \section{Exact Solutions to the Mode Equation}

Eq.~(\ref{ff}) can be solved exactly for the profile
\be \label{p} c = (A+B\tanh kz)^{-1/2} \;, \ee
with $A$ and $B$ positive constants.
In order for this profile to describe the exterior of an acoustic black hole we define the asymptotic values at, respectively, $z\to -\infty$ and $z\to +\infty$ as
\be \label{bc} (A-B)^{-1/2}=v_0, \ (A+B)^{-1/2}=c_R\ . \ee
As can be verified by substitution into Eq.~\eqref{tr}
\be \label{x} x=\frac{1}{2k}(1-\frac{v_0^2}{c_R^2})\ln(e^{2z}+1)\ ,\ee
with the result that
\be \label{cext} c(x)=\frac{c_R}{\sqrt{1+(\frac{c_R^2}{v_0^2}-1)e^{-\frac{2c_R^2kx}{c_R^2-v_0^2}}}}\ . \ee
Thus with the choices~\eqref{bc} the profile gives $c = c_R$ at $x=+\infty$ and $c=v_0$ at the horizon $x=0$.  The horizon's surface gravity is \be \label{sg} \kappa=\frac{dc}{dx}\vline_{_{\rm hor}}=kv_0 \ .\ee

To solve Eq.\ (\ref{ff}) with the profile given by Eq.\ (\ref{p}), we introduce the notation
\be \label{z} y_{\pm}\equiv \frac{1}{2}(1\mp\tanh kz) \;, \ee
and write down the two linear independent solutions as
\be \label{mm} \varphi_{\pm}=y_{\pm}^{-\frac{i\omega\sqrt{A\pm B}}{2k}}(1-y_{\pm})^{\frac{i\omega\sqrt{A\mp B}}{2k}} F_{\pm}(a_{\pm}, b_{\pm} ,c_{\pm} ; y_{\pm})\;, \ee
where $F$ is the hypergeometric function with
\bea a_{\pm} &=&1-\frac{i\omega}{2k}(\sqrt{A\pm B}-\sqrt{A\mp B})\ ,\nonumber \\ b_{\pm} &=& -\frac{i\omega}{2k}(\sqrt{A\pm B}-\sqrt{A\mp B})\ , \nonumber \\ c_{\pm} &=& 1-\frac{i\omega}{k} \sqrt{A\pm B}\ . \label{zz} \eea

 From $\varphi_+$ we can construct the normalized solution $\varphi_H$.
 First note that Eq. (\ref{z}) implies that $y_+=0$ at infinity ($z=+\infty$) and $y_+=1$ at the horizon ($z=-\infty$).
To uncover the behavior of $\varphi_+$ on the horizon we make use of the following identity
\bea \label{rel} && F_+(a_+,b_+,c_+ ; y_+) =  \frac{\Gamma(c_+)\Gamma(c_+ -a_+ -b_+)}{\Gamma(c_+ -a_+)\Gamma(c_+ -b_+)}F_+(a_+,b_+, a_+ +b_+ -c_+ +1;1-y_+)  \\ && + (1-y_+)^{c_+ -a_+ -b_+}\frac{\Gamma(c_+)\Gamma(a_+ +b_+ -c_+)}{\Gamma(a_+)\Gamma(b_+)}F_+(c_+ -a_+,c_+ -b_+, c_+ -a_+ -b_+ +1; 1-y_+)\ . \nonumber \eea
Inserting this for $F_+(a_+,b_+,c_+;y_+)$ in Eq.\ (\ref{mm}),  making use of the fact that $F(a_+,b_+,c_+ ; 0)=1$, and noting that $1-y_+ \approx e^{2 k z}$, it can be seen that
 near the horizon $\varphi_+$  consists of two terms; the first (multiplied as usual by
$e^{-i\omega t_S}$) describes an outgoing plane wave $e^{i\frac{\omega}{v_0}z}$ while the second describes an ingoing plane wave $e^{-i\frac{\omega}{v_0}z}$.
Comparison with~\eqref{uu} gives
\be \label{ch} \varphi_H=\sqrt{\frac{v_0}{4\pi\omega}}\ \frac{\Gamma(c_+ -a_+)\Gamma(c_+ -b_+)}{\Gamma(c_+)\Gamma(c_+ -a_+ -b_+)}\ \varphi_+ \;, \ee
and
\bea \label{rh} R_H &=& \frac{\Gamma(a_+ +b_+ -c_+)\Gamma(c_+ -a_+)\Gamma(c_+ -b_+)}{\Gamma(a_+)\Gamma(b_+)\Gamma(c_+ -a_+ -b_+)}
\nonumber \\ &=& \frac{\Gamma(\frac{i\omega}{kv_0})}{\Gamma(-\frac{i\omega}{kv_0})}\frac{ \Gamma(-\frac{i\omega}{2k}(\frac{1}{v_0}+\frac{1}{c_R}))\Gamma(1-\frac{i\omega}{2k}(\frac{1}{v_0}+\frac{1}{c_R})) }
{ \Gamma(-\frac{i\omega}{2k}(\frac{1}{c_R}-\frac{1}{v_0}))\Gamma(1-\frac{i\omega}{2k}(\frac{1}{c_R}-\frac{1}{v_0})) }\ . \eea

 For large $z$,  $y_+ \approx e^{-2 k z}$.  Comparison of~\eqref{ch} and~\eqref{mm} for large $z$ with
 the lower expression of~(\ref{uu}) gives
\bea \label{th} T_H &=& \sqrt{\frac{v_0}{c_R}} \frac{\Gamma(c_+ -a_+)\Gamma(c_+ -b_+)}{\Gamma(c_+)\Gamma(c_+ -a_+ -b_+)}
\nonumber \\ &=& \sqrt{\frac{v_0}{c_R}}\ \frac{\Gamma(-\frac{i\omega}{2k}(\frac{1}{v_0}+\frac{1}{c_R}))\Gamma(1-\frac{i\omega}{2k}(\frac{1}{v_0}+\frac{1}{c_R}))}{\Gamma(1-\frac{i\omega}{kc_R})\Gamma(-\frac{i\omega}{kv_0})} \ . \eea
The scattering coefficients satisfy the unitarity relation
\be \label{uru} |T_H|^2 + |R_H|^2 =1 \;.\ee
In the $\omega\to 0$ limit
\be \label{so} T_H\sim \frac{2\sqrt{v_0c_R}}{c_R+v_0}\ \ \ \ , R_H\sim \frac{ c_R-v_0}{c_R+v_0}\  \ee
as predicted in Refs.~\cite{abfp2,afb} while for large $\omega$, $T_H$ is a phase and $R_H$ vanishes.  Thus as expected, in the high frequency limit, we have complete transmission.
The gray-body factor is
\be \label{gf} \Gamma=|T_H|^2= \frac{\sinh \frac{\pi \omega}{kv_0} \sinh \frac{\pi \omega}{kc_R}}{\sinh^2 \left[\frac{\pi \omega}{2k}(\frac{1}{v_0}+\frac{1}{c_R})\right]}\ .
\ee
As already stressed in Refs.~\cite{abfp2,afb}, the fact that the gray-body factor goes to a nonzero constant in the low frequency limit implies that the analog Hawking radiation of 1D BEC acoustic black holes is dominated by an infinite number ($\frac{1}{\omega}$) of soft phonons.  From an experimental point of view a direct detection of the emitted phonons is difficult and one infers their features from density correlation measurements (see for instance \cite{jeff}).
In the gravitational case, a similar feature is present in the Hawking radiation from Schwarzschild-de Sitter black holes. The behavior is very different for Schwarzschild black holes where $\Gamma \sim \omega^2$ as $\omega^2\to 0$, so that the low frequency emission is absent.

The construction of $\chi_I$ proceeds in a similar fashion starting from $\varphi_-$. In this case $y_-=1$ at infinity and $y_-=0$ at the horizon. Using the identity (\ref{rel}) (with $F_+, a_+, b_+, c_+, y_+ \to F_-,\ a_-, b_-, c_-, y_-$) one finds
\be \label{ci} \chi_I=\sqrt{\frac{c_R}{4\pi\omega}}\  \frac{\Gamma(c_- -a_-)\Gamma(c_- -b_-)}{\Gamma(c_-)\Gamma(c_- -a_- -b_-)}\ \varphi_-  \;. \ee
Comparison of~\eqref{ci} and~\eqref{mm} with~\eqref{II} gives
\bea \label{ri} R_I &=& \frac{\Gamma(a_- +b_- -c_-)\Gamma(c_- -a_-)\Gamma(c_- -b_-)}{\Gamma(a_-)\Gamma(b_-)\Gamma(c_- -a_- -b_-)}
\nonumber \\ &=& \frac{\Gamma(\frac{i\omega}{kc_R})}{\Gamma(-\frac{i\omega}{kc_R})}\frac{ \Gamma(-\frac{i\omega}{2k}(\frac{1}{v_0}+\frac{1}{c_R}))\Gamma(1-\frac{i\omega}{2k}(\frac{1}{v_0}+\frac{1}{c_R})) }{ \Gamma(-\frac{i\omega}{2k}(\frac{1}{v_0}-\frac{1}{c_R}))\Gamma(1-\frac{i\omega}{2k}(\frac{1}{v_0}-\frac{1}{c_R})) }\ , \eea
and
\bea \label{ti} T_I &=& \sqrt{\frac{c_R}{v_0}} \frac{\Gamma(c_- -a_-)\Gamma(c_- -b_-)}{\Gamma(c_-)\Gamma(c_- -a_- -b_-)}
\nonumber \\ &=& \sqrt{\frac{c_R}{v_0}}\ \frac{\Gamma(-\frac{i\omega}{2k}(\frac{1}{v_0}+\frac{1}{c_R}))\Gamma(1-\frac{i\omega}{2k}(\frac{1}{v_0}+\frac{1}{c_R}))}{\Gamma(1-\frac{i\omega}{kv_0})\Gamma(-\frac{i\omega}{kc_R})}=T_H \ . \eea
These coefficients satisfy the unitarity relation $|T_I|^2 + |R_I|^2 =1$. In the limit $\omega  \to 0$ we have
$T_I=T_H \sim \frac{2\sqrt{v_0c_R}}{c_R+v_0}\ , R_I\sim \frac{ v_0-c_R}{c_R+v_0}$.  In the large $\omega$ limit, $|T_I|\rightarrow 1$ and $R_I \rightarrow 0$ as expected.

Finally, the coefficients in (\ref{rh}), (\ref{th}), (\ref{ri}), (\ref{ti}) satisfy $R_H^* T_I + T_H^* R_I=0$.

\section{Summary and Conclusions}

We have found a complete set of exact solutions to the mode equation in the region outside of the acoustic horizon for a 1D BEC analog black hole with
constant speed $v_0$ and density $n$ and with
the speed of sound profile~\eqref{cext}.  From these we have obtained analytic expressions for the scattering coefficients and gray-body factor.  The speed of sound profile is a well behaved one which asymptotically approaches a
constant at infinity and is equal to the flow speed at the horizon.    In principle such a profile could be set up in the laboratory.\footnote{Note that our results apply also to more general models, in which $v,\ n,\ c$ are nontrivial. In this case, we still get an equation like (\ref{ff}) with the substitution $\frac{1}{c^2}\to \frac{n^2}{c^2}$, and the coordinate $z$ is defined via $dz=\frac{dx}{n(1-\frac{v^2}{c^2})}$ instead of (\ref{tr}). }

The fact that we have analytic expressions for the scattering coefficients and the gray-body factor has allowed us to verify the predictions made for these quantities in the low
frequency limit in~\cite{abfp2,afb}.  Two important subtleties in the derivations of those predictions involved the order in which the limits $\omega \rightarrow 0$ and $x \rightarrow 0$ were taken and the order in which the limits $\omega \rightarrow 0$ and $x \rightarrow \infty$ were taken.
No such subtlety occurs here as the scattering coefficients have been computed analytically for all values of $\omega$.

\begin{acknowledgments}  This work was supported in part by the National Science Foundation under Grant Nos.  PHY-1308325 and PHY-1505875.
\end{acknowledgments}

{}
\end{document}